\documentclass[10pt,
               aps,
               prl,
               twocolumn,
               showpacs, 
               groupedaddress]{revtex4-2}
\usepackage[utf8]{inputenc}
\usepackage{graphicx}
\graphicspath{{images/}}
\usepackage[caption=false]{subfig}
\usepackage{amsmath, amsfonts, amssymb}
\usepackage{hyperref}
\usepackage{tikz}
\makeatletter
\newcommand*{\rom}[1]{\expandafter\romannumeral #1}
\makeatother

\begin{document}

\title{Inertial active harmonic particle with memory escape induced by viscoelastic suspension}

\author{F Adersh}

\affiliation{Department of Physics, University of Kerala, Kariavattom, Thiruvananthapuram-$695581$, India}

\author{M Muhsin}

\affiliation{Department of Physics, University of Kerala, Kariavattom, Thiruvananthapuram-$695581$, India}

\author{M Sahoo}
\email{jolly.iopb@gmail.com}
\affiliation{Department of Physics, University of Kerala, Kariavattom, Thiruvananthapuram-$695581$, India}

\date{\today}

\begin{abstract}
We investigate the self-propulsion of an inertial active particle confined in a two-dimensional harmonic trap. The particle is suspended in a non-Newtonian or viscoelastic suspension with a friction kernel that decays exponentially with a time constant characterizing the memory timescale or transient elasticity of the medium. By solving the associated non-Markovian dynamics, we identify two regimes in parameter space distinguishing the oscillatory and non-oscillatory behavior of the particle motion. By simulating the particle trajectories and exactly calculating the steady state probability distribution functions and mean square displacement, interestingly, we observe that with an increase in the memory time scale, the elastic bound of suspension dominates over the influence of harmonic trap. As a consequence, the particle can escape out of the trap without approaching steady state. On the other hand, with an increase in the duration of the activity, the particle becomes trapped by the harmonic confinement. 
\end{abstract}

\maketitle
\section{INTRODUCTION}\label{sec:intro}

The physics of active matter represents a rapidly advancing field of research that has gained substantial interest across various scientific disciplines~\cite{bechinger2016active, Ramaswamy2017active, Gompper2020roadmap, pietzonka2021oddity, magistris2015intro}, particularly in the realm of biophysics and bioengineering. Active or self-propelling systems constitute entities that are driven out of equilibrium by harnessing energy from their environment to generate directed motion. Active matter primarily includes microorganisms like E.coli, ciliates, and other motile bacteria, but they can also be synthesized artificially. Janus particles~\cite{walther2013janus, howse2007self} and microrobots~\cite{wang2021emergent} are some of the prominent examples of artificially synthesized active matter. They can mimic biological motility and find applications in many emerging scientific and technological domains like material science and nanotechnology. 
Some of the notable theoretical models for exploring active matter include active Ornstein-Uhlenbeck particle (AOUP) model\cite{lehle2018analyzing, bonilla2019active, martin2021statistical, caprini2019active, caprini2021inertial, dabelow2019irreversibility, berthier2019glassy, wittman2018effective, fily2019self, mandal2017entropy, fodor2016far, muhsin2021orbital}, Active Brownian particle model (ABP)~\cite{hagen2009NonGaussian, hagen2011brownian, cates2013when, kanaya2020steady, buttinoni2013dynamical, fily2012athermal, stenhammar2014phase, bialke2015negative, solon2015pressure, caprini2021collective, caprini2020hidden, roon2022role, scholz2018inertial, mandal2019motility}, run and tumble particle (RTP) model~\cite{berg1972chemo, bartens2012probability}, etc. The AOUP model introduces activity into the system through a stochastic force that follows the Ornstein-Uhlenbeck process. The ABP model considers both translational and rotational diffusion of particles, while RTPs exhibit a ``run" phase characterized by ballistic motion with constant speed followed by periodic tumbling.

An inertial active particle while self-propelling within a harmonic confinement can exhibit oscillatory motion and the frequency of oscillation depends on the inertia and strength of the harmonic confinement ~\cite{nguyen_active_2022}. Such an active harmonic particle while self-propelling in a two-dimensional (2d) plane, the oscillatory motion turns in the performance of rotational trajectories ~\cite{caprini2021inertial, arsha2021velocity, muhsin2022inertial}. Further, an active particle can also exhibit oscillation in its motion while suspended in a viscoelastic environment~\cite{paraan2008brownian, sevilla2019generalized}. Such induced oscillations often serve as distinctive characteristic of non-equilibrium particle activity~\cite{solano2022nonequilibrium}. Moreover, such a viscoelastic particle while confined in a plane, it performs circular motion~\cite{narinder2018memory}. 

In this paper, we delve into the dynamics of an inertial active Ornstein-Uhlenbeck particle (AOUP) suspended in a non-Newtonian bath and confined within a harmonic potential. The viscoelastic memory of the environment influences the particle motion in both qualitative and quantitative ways. Examples include the memory induced delay between the effective self-propulsion force and particle orientation in an overdamped active particle \cite{sprenger2022active}, strong enhancement of directional changes in the periodically modulated Run and Tumble particles in viscoelastic bath \cite{lozano2018run}, effective repulsion among active particles induced by viscoelastic fluid while moving near a boundary\cite{narinder2019active}, etc. 
By simulating and exactly solving our present model dynamics, we observe that for specific choice of parameters, the particle exhibits oscillatory motion and the frequency of oscillation exhibits a non-monotonic dependence on both inertial and harmonic time scales. The simulated particle trajectories confirm that the particle adopts rotational motion in the oscillatory regime of parameter space. Moreover, an increase in the memory timescale of the medium results in the effective dominance of elastic dissipation over harmonic confinement. As a consequence, the particle can escape out of the potential or blow off without approaching a steady state. On the other hand, an increase in the activity time results in trapping of the particle by the harmonic confinement. These findings are consistent with the analytically computed steady state results for probability distribution functions and mean square displacement.

\section{MODEL}\label{sec:model}
We consider the 2d motion of an inertial active Ornstein-Uhlenbeck particle in a viscoelastic environment. The particle is confined in a harmonic potential $U(x,y) = \frac{1}{2} k (x^2 + y^2)$ with $k$ as the harmonic constant. The dynamics is non-Markovian due to the non-Newtonian nature of the surrounding medium and hence can be described using the generalized Langevin's equation of motion \cite{sevilla2019generalized, muhsin2021orbital}
\begin{equation}
   m\ddot{\bf r}(t)=-\gamma\int_{0}^{t}\lambda(t-t') \dot{\bf r}(t')\, dt' - \boldsymbol{\nabla} U + \boldsymbol{\xi}(t).
     \label{eq:maineqmotion-vector}
\end{equation}

Here, ${\bf r}(t) = x(t) \hat{i} + y(t) \hat{j}$ is the position vector of the particle in the $x-y$ plane and $m$ represents the mass of the particle. 
The first term in RHS defines the viscous drag and is characterized by an exponentially decaying memory kernel $\lambda(\tau)$ of the form
\begin{equation}
\lambda\left(\tau\right)=\begin{cases}
\frac{1}{t_c'}e^{-\frac{\tau}{t_c'}}; &  \tau\geq 0,\\
0~~~~~~~~~; & \tau < 0.
\end{cases}
\end{equation}
The above memory kernel represents that the medium exhibits a transient elasticity that exponentially decays to viscous behavior with a timescale $t_c'$. Hence, $t_c'$ can be regarded as the memory time scale or elastic dissipation timescale of the system. The term $\boldsymbol{\xi}(t)$ in Eq.~\eqref{eq:maineqmotion-vector} represents the athermal noise associated with the dynamics which follows the Ornstein-Uhlenbeck (OU) process
\begin{equation}
t_c\dot{\boldsymbol{\xi}}(t) = -\boldsymbol{\xi}(t) + \sqrt{D} \boldsymbol{\eta}(t).
\label{eq:noise}
\end{equation}
Here, $t_c$ represents the self-propulsion or activity timescale of the dynamics, $D$ being the strength of noise and $\boldsymbol{\eta}(t)$ is the delta-correlated white noise. $\boldsymbol{\xi}(t)$ has the properties

\begin{equation}
\langle \xi_\alpha(t) \rangle =0, \qquad  \langle \xi_\alpha(t)\xi_\beta(t') \rangle = \frac{\delta_{\alpha\beta}}{2t_c} e^{-\frac{|t-t'|}{t_c}}.
\label{eq:noise_stat}
\end{equation}
When $t_c = t_c'$, the generalized fluctuation-dissipation theorem is validated and the system approaches thermal equilibrium for $D = 2 \gamma k_BT$.

Now we define the physical quantities of interest. The mean displacement (MD) of the particle from the initial position is given by
\begin{equation}
    \langle \Delta{\bf r}(t) \rangle = \langle {\bf r}(t) - {\bf r}(0) \rangle.
    \label{eq:md}
\end{equation}
Similarly, the mean square displacement (MSD) can be obtained as
\begin{equation}
\begin{split}
    \langle {\Delta{\bf r}(t)}^2 \rangle &= \langle \left({\bf r}(t) - {\bf r}(0)\right)^2 \rangle.
    \label{eq:msd}
\end{split} 
\end{equation}

To obtain the particle trajectory, we perform a numerical simulation of Langevin dynamics [Eq.~\eqref{eq:maineqmotion-vector}]. The integration of the equation of motion [Eq.~\eqref{eq:maineqmotion-vector}] is carried out using a second-order modified Euler method with a timestep of $10^{-3}$. The OU noise is realized using the Fox algorithm\cite{fox1988fast}. In the next section, we discuss the results obtained from the numerical simulation and analytical calculations.    

\section{RESULTS AND DISCUSSION}\label{sec:result}
\begin{figure}
    \centering
    \includegraphics[width=\linewidth]{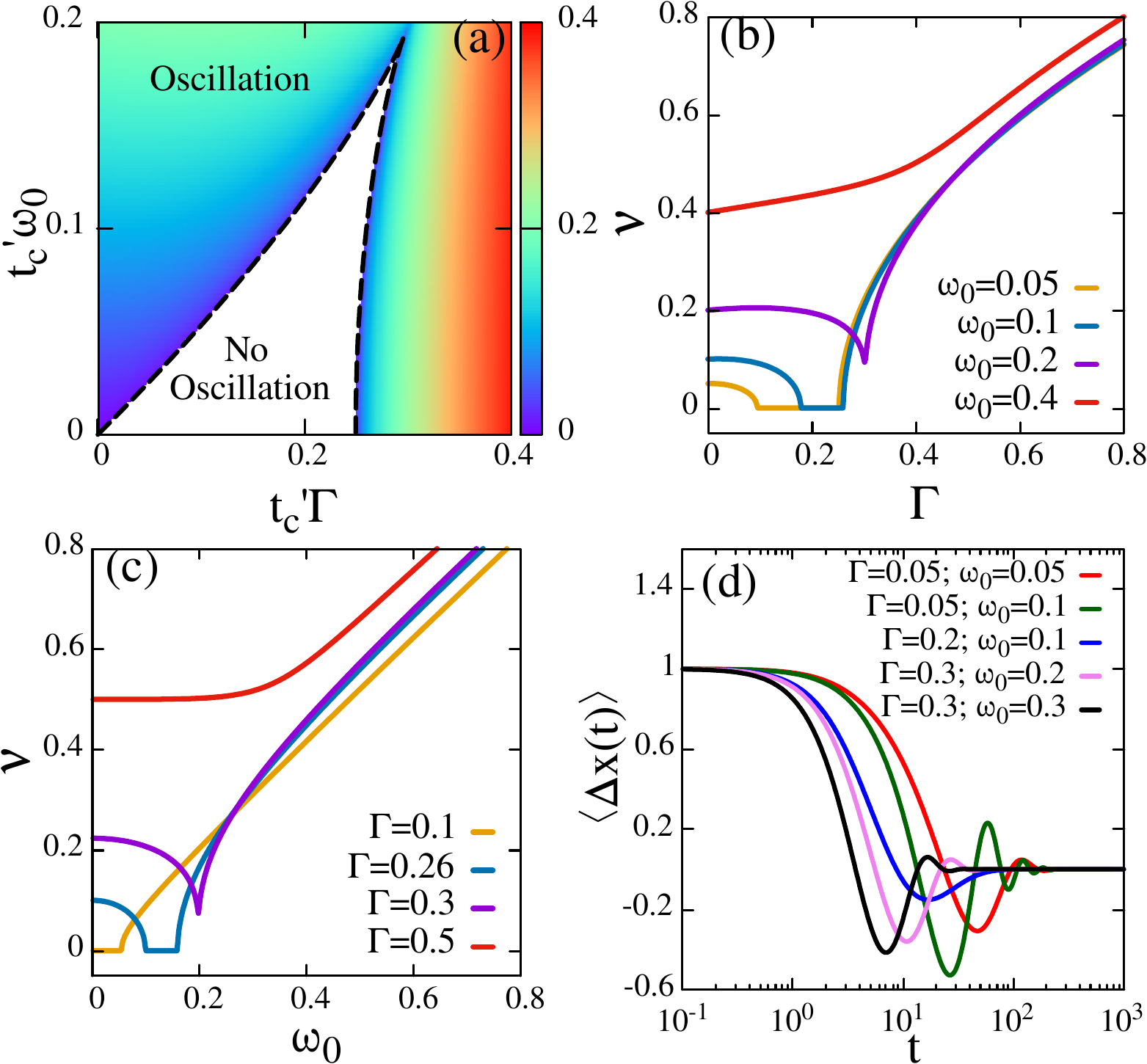}
    \caption {(a) Phase diagram separating the oscillatory and non-oscillatory regimes in the $t_c' \Gamma-t_c' \omega_{0}$ parameter space [Eq.~\eqref{eq:soln_HT}]. The color map in (a) shows the evolution of frequency of oscillation ($\nu$) with $t_c' \Gamma$ and $t_c' \omega_{0}$. (b) $\nu$ vs $\Gamma$ for different values of $\omega_0$ and for a fixed $t_c'=1$. (c) $\nu$ vs $\omega_0$ for different values of $\Gamma$ and for a fixed $t_c'=1$. (d) The time evolution of the x-component of mean displacement (MD) $\langle \Delta x(t) \rangle$ for different values of $\omega_0$ and $\Gamma$ with $t_c'=1.0$. The other common parameters are $m = t_c = D = 1$.}.
    \label{fig:MD_HT_nomag}
\end{figure}

Solving the dynamics [Eq.~\eqref{eq:maineqmotion-vector}], with initial conditions ${\bf r}(0) = {\bf r}_0$ and ${\bf \dot{r}}(0) = {\bf v}_0$, the solution can be obtained as
\begin{equation}
{\bf r(t)} = \sum_{i=1}^{3} a_i \biggr[e^{s_i t}\left({\bf v_0} +s_i {\bf r_0}\right)+\int_{0}^{t} e^{s_i(t-t')}\xi(t') \,dt'\biggr].
\label{eq:soln_HT}
\end{equation}
Here, $s_i$'s are the roots of the equation
\begin{equation}
    s^3+\frac{s^2}{t_c'}+s \left(\frac{\Gamma }{ t_c'}+\omega _0^2\right)+\frac{\omega _0^2}{t_c'}=0,
    \label{eq:cubic}
\end{equation}
which are given by
\begin{equation}
    s_i = \frac{1}{3t_c'}\left[ -1 - \frac{\omega^{i-1} \Delta_0}{\sqrt[3]{\Delta_1}} + \omega^{i-1} \sqrt[3]{\Delta_1} \right], \quad i \in \{1,2,3\}
    \label{eq:si-value}
\end{equation}
with
\begin{equation}
\begin{split}
    \Delta_0 &= -1 + 3\left[ t_c'\Gamma + (t_c'\omega_0)^2 \right], \\
    \Delta_1 &= \frac{-2 + 9 t_c'\Gamma - 18(t_c'\omega_0)^2 + \sqrt{-\Delta}}{2}, \\
    \Delta &= -\left[2 - 9 t_c'\Gamma + 18(t_c'\omega_0)^2\right]^2 - 4\Delta_0^3, \quad \text{and} \\
    \omega &= \frac{-1 + j \sqrt{3}}{2}.
\end{split}
\label{eq:Delta}
\end{equation}
Here, $\Gamma = \frac{\gamma}{m}$, $\omega_0 = \sqrt{\frac{k}{m}}$ and $j = \sqrt{-1}$. The coefficients $a_i$'s in Eq.~\eqref{eq:soln_HT} are given by
\begin{equation}
    \begin{split}
        &a_1=\frac{s_1 t_c'+1}{\left(s_1-s_2\right) \left(s_1-s_3\right) t_c'},~
        a_2=\frac{s_2 t_c'+1}{\left(s_2-s_1\right) \left(s_2-s_3\right) t_c'}\\
        &\text{and}~~ a_3=\frac{s_3 t_c'+1}{\left(s_2-s_3\right) \left(s_1-s_3\right) t_c'}.
    \end{split}
\end{equation}

Solving Eq.~\eqref{eq:cubic}, it is realized that the roots of Eq.~\eqref{eq:cubic} can be real as well as complex. Positive and negative values of $\Delta$ result in some of the roots of Eq.~\eqref{eq:cubic} to be complex.
The complex roots are responsible for the oscillatory solution of the dynamics [Eq.~\eqref{eq:soln_HT}], whereas the real roots provide an  
exponentially decaying solution without any oscillatory behavior. Thus, there exist two separate regimes associated with the oscillatory and non-oscillatory behavior of the solution. Since the roots $s_i$ are independent of $t_c$, these oscillatory and non-oscillatory regimes are presented in a phase diagram of $t_c' \omega_{0}$-$t_c'\Gamma$ parameter space [see Fig.~\ref{fig:MD_HT_nomag}(a)]. The boundary separating these two regimes in $t_c'\omega_0$~--~$t_c'\Gamma$ space is given by $\Delta = 0$. The color map in Fig.~\ref{fig:MD_HT_nomag}(a) shows the evolution of frequency of oscillation ($\nu$) with both $\omega_{0}$ and $\Gamma$ for a fixed $t_c'$ value. The frequency of oscillation is computed as the imaginary part of $s_{i}$ (Eq.~\eqref{eq:si-value}). In Fig.~\ref{fig:MD_HT_nomag}(b), we have shown the variation of $\nu$ with $\Gamma$ for different values of $\omega_0$ by keeping $t_c'$ fixed. For small values of $\omega_0$, $\nu$ as a function of $\Gamma$ initially decreases, becomes zero and shows a plateau for the range of $\Gamma$ that falls in the non-oscillatory regime of Fig.~\ref{fig:MD_HT_nomag}(a) and finally increases with increase in $\Gamma$ value. For higher values of $\omega_0$, this non-monotonic behavior of $\nu$ with $\Gamma$ disappears and it monotonically increases with $\Gamma$. Similarly, Fig.~\ref{fig:MD_HT_nomag}(c) shows the variation of $\nu$ with $\omega_0$ for different values of $\Gamma$ and for a fixed $t_c'$. For lower $\Gamma$ values, $\nu$ as a function of $\omega_{0}$ shows a non-monotonic behavior, it decreases, approaches zero, and finally increases with $\omega_0$. 
For higher $\Gamma$ values, $\nu$ becomes an increasing function of $\omega_0$. 
For further exploration of these two regimes, in Fig.~\ref{fig:MD_HT_nomag}(d), we plot the x-component of MD $\langle \Delta x(t) \rangle$ for different values of $\Gamma$ and $\omega_0$ keeping $t_c' = 1$.  
Using Eq.~\eqref{eq:soln_HT}, MD [Eq.~\eqref{eq:md}] of the particle can be calculated as
\begin{equation}
     \langle \Delta{\bf r}(t) \rangle =\sum_{i=1}^{3} a_i e^{s_i t}  \Bigr[{\bf v_0}+{\bf r_0}s_i \Bigr] - {\bf r_0}.
    \label{eq:MD-HT-no-mag}
\end{equation}
In the lower powers of $t$, it can be expanded as
\begin{equation}
    \langle \Delta{\bf r}(t) \rangle = -{\bf r_0} + {\bf v_0} t-\frac{{\bf r_0}\left(\Gamma +4 t_c'\omega_0^2\right)}{8 t_c'}t^2 +O\left(t^3\right).
    \label{eq:mdtransient_free_nomagss}
\end{equation}
From Fig.~\ref{fig:MD_HT_nomag}(d), it is to be noted that $\langle \Delta x(t) \rangle$ does not show any oscillatory behavior for the parameters that lie in the no-oscillation regimes of $t_c' \Gamma$-$t_c'\omega_0$ parameter space (solid blue curve for $\Gamma = 0.2$ and $\omega_0 = 0.1$). On the other hand, $\langle \Delta x(t) \rangle$ exhibits an intermediate time oscillatory behavior(curves other than blue in color) for the parameters that fall in the oscillatory regime of $t_c' \Gamma$-$t_c'\omega_0$ parameter space. 
\begin{figure*}[!ht]
    \centering
  \includegraphics[width=0.7\linewidth]{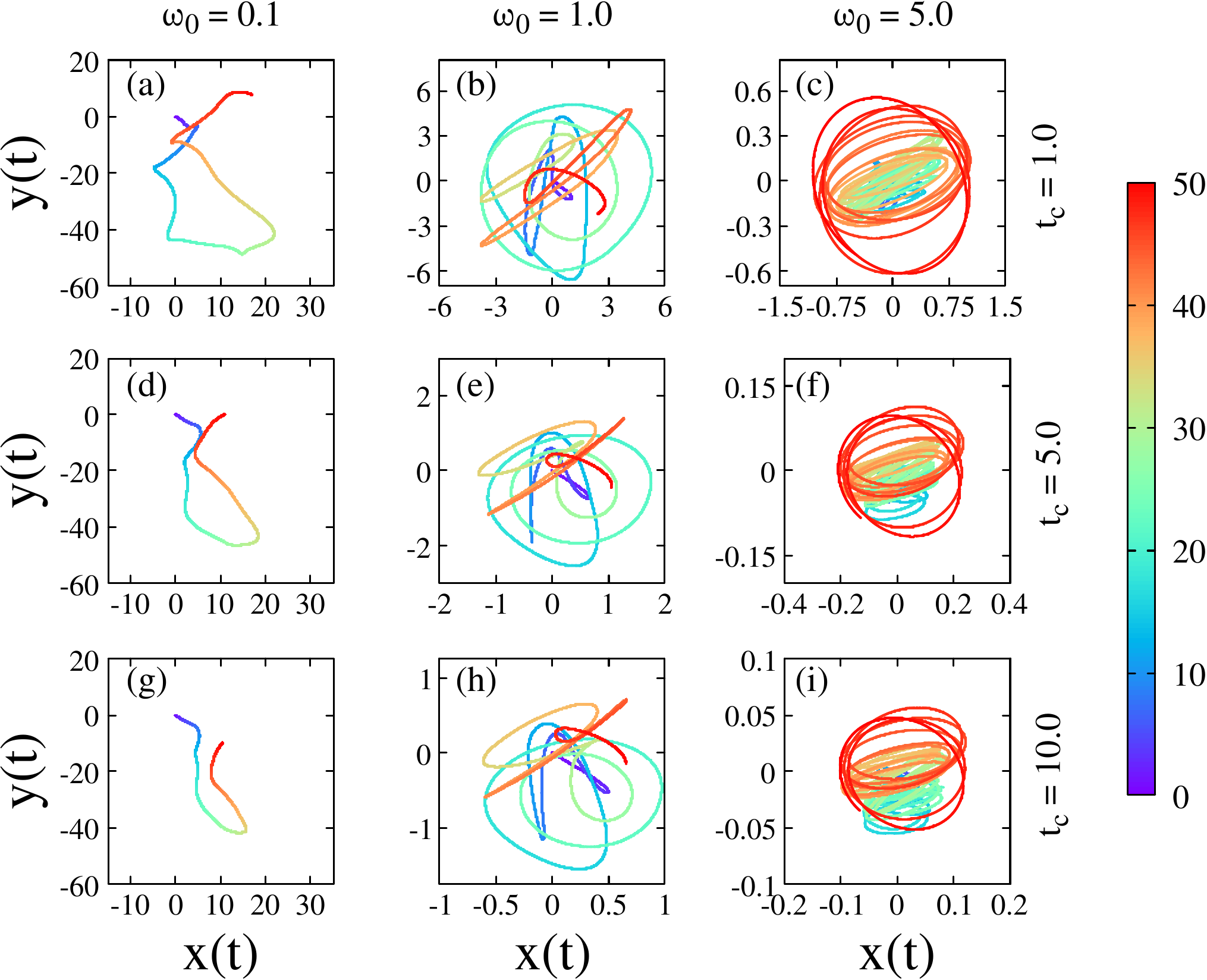}
    \caption{The simulated particle trajectories in the $xy$-plane for different values of $\omega_0$ are plotted for $t_c=1$ in (a),(b) and (c), for $t_c=5$ in (d), (e) and (f),  and for $t_c=10$ in (g), (h) and (i), respectively. The other common parameters are $m = t_c' = D = 1$, and $\Gamma=0.2$. The color map shows the time evolution of the trajectories.}
    \label{fig:Traj_g10_HT_nomag_tc}
\end{figure*}

In Fig.~\ref{fig:Traj_g10_HT_nomag_tc}, we plot the instantaneous 2d particle trajectories for different values of $\omega_0$ and $t_c$. Each column of Fig.~\ref{fig:Traj_g10_HT_nomag_tc} corresponds to a fixed value $\omega_0$ and each row corresponds to a fixed value of $t_{c}$. For a given set of parameters, with an increase in $\omega_0$ value from left to right in any of the rows, the particle passes from no-oscillatory to oscillatory regime. The trajectories reflect random self-propulsion for the parameters that lie in the no-oscillatory regimes [see Figs.~\ref{fig:Traj_g10_HT_nomag_tc}(a), (d) and (g)]. However, the particle performs rotational trajectories for the parameters that lie in the oscillatory regime of parameter space [see Figs.~\ref{fig:Traj_g10_HT_nomag_tc}(b), (c), (e), (f), (h) and (i)]. Further, the trajectories show stronger confinement of the particle with increase in $\omega_0$ value as expected. Similarly, in any of the columns, the trajectories get suppressed with increase in $t_c$ value. This observation suggests the trapping or confinement of the particle around the centre of the potential with increase in persistent duration of activity.  
\begin{figure*}
    \centering
  \includegraphics[width=0.7\linewidth]{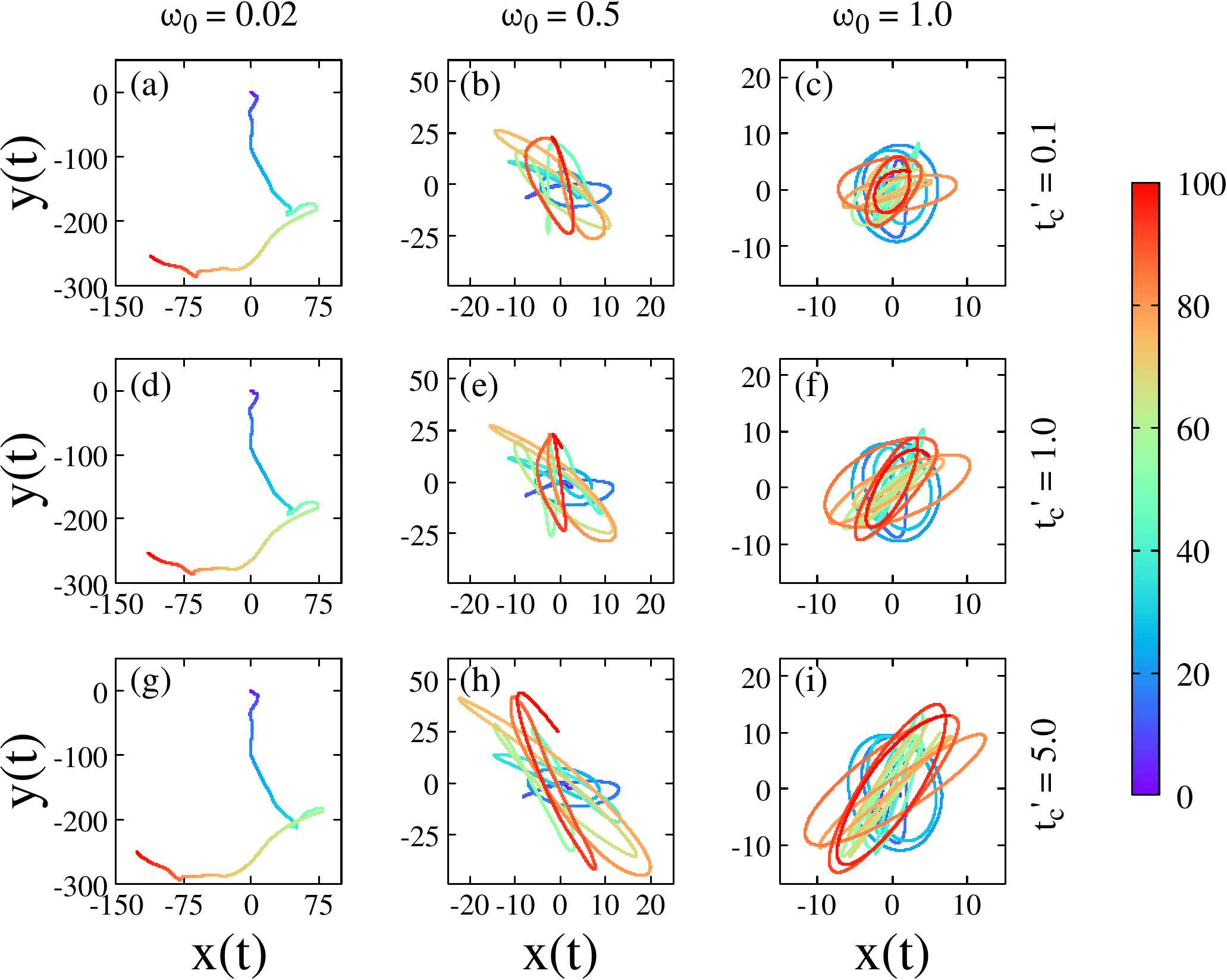}
    \caption{The simulated particle trajectories in the $xy$-plane for different values of $\omega_0$ are plotted for $t_c'=0.1$ in (a), (b) and (c), for $t_c'=1$ in (d), (e) and (f),  and for $t_c'=5.0$ in (g), (h) and (i), respectively. The other common parameters are $m = 1$, $\Gamma=0.05$, $t_c=1.0$ and $D = 1.0$. The color map shows the time evolution of the trajectories.}
    \label{fig:Traj_g10_HT_nomag_tcp}
\end{figure*}

Similarly, the 2d particle trajectories for different values of $\omega_0$ and $t_c'$ are shown in Fig.~\ref{fig:Traj_g10_HT_nomag_tcp}, with $\omega_{0}$ increasing along a row from left to right and $t_c'$ increasing along a column from top to bottom. Similar to the observations in Fig.~\ref{fig:Traj_g10_HT_nomag_tc}, for a fixed $t_c'$ value, with increase in $\omega_{0}$ along a row, the parameters are chosen such that the particle passes from no-oscillatory to oscillatory regime. The particle performs random self-propulsion in the no-oscillatory regime, i.e., for $\Delta > 0$ [see Fig.~\ref{fig:Traj_g10_HT_nomag_tcp}(a), (d) and (g)]. For higher $\omega_{0}$ values, the particle enters the oscillatory regime (since $\Delta$ becomes negative) and makes a transition from its random activity to twisting or rotational motion [see Figs.~\ref{fig:Traj_g10_HT_nomag_tcp}(b), (c), (e), (f), (h) and (i)]. Further, with increase in $t_c'$ value from top to bottom in a column, the trajectories get enhanced. At the same time, the trajectories become more asymmetric and take almost elliptical shape. The enhancement of trajectory as an increasing function of $t_c'$ can be attributed to the fact that for high value of $t_c'$, the system retains memory for a prolonged period, resulting in less frequent change in the direction of the particle velocity. As a consequence, the particle takes longer time to complete the orbit.

Next, we have calculated the MSD [Eq.~\eqref{eq:msd}] using Eq.~\eqref{eq:soln_HT} as
\begin{widetext}
\begin{equation}
\begin{split}
     \langle \Delta{\bf r}^2(t) \rangle = &\langle \left({\bf r}(t) - {\bf r}(0)\right)^2 \rangle= \left( \langle \Delta{\bf r}(t) \rangle\right)^2  +\sum_{i=1}^3\sum_{j=1}^3 \frac{a_i a_j^* D}{m^2} \Biggl[ \frac{t_c e^{t \left(s^*_j-\frac{1}{t_c}\right)}}{\left(t_c s_i+1\right) \left(1-t_c s^*_j\right)}+\frac{t_c e^{t
   \left(s_i-\frac{1}{t_c}\right)}}{\left(1-t_c s_i\right) \left(t_c s^*_j+1\right)} \\
   & +\frac{t_c \left(s_i+s^*_j\right)-2}{\left(s_i+s^*_j\right) \left(t_c s_i-1\right) \left(t_c s^*_j-1\right)}+\frac{e^{t \left(s_i+s^*_j\right)}
   \left(t_c \left(s_i+s^*_j\right)+2\right)}{\left(s_i+s^*_j\right) \left(t_c s_i+1\right) \left(t_c s^*_j+1\right)} \Biggr].
      \label{eq:MSD-HT-no-mag}
\end{split}
\end{equation}
\end{widetext}

\begin{figure}
    \centering
    \includegraphics[scale=0.35]{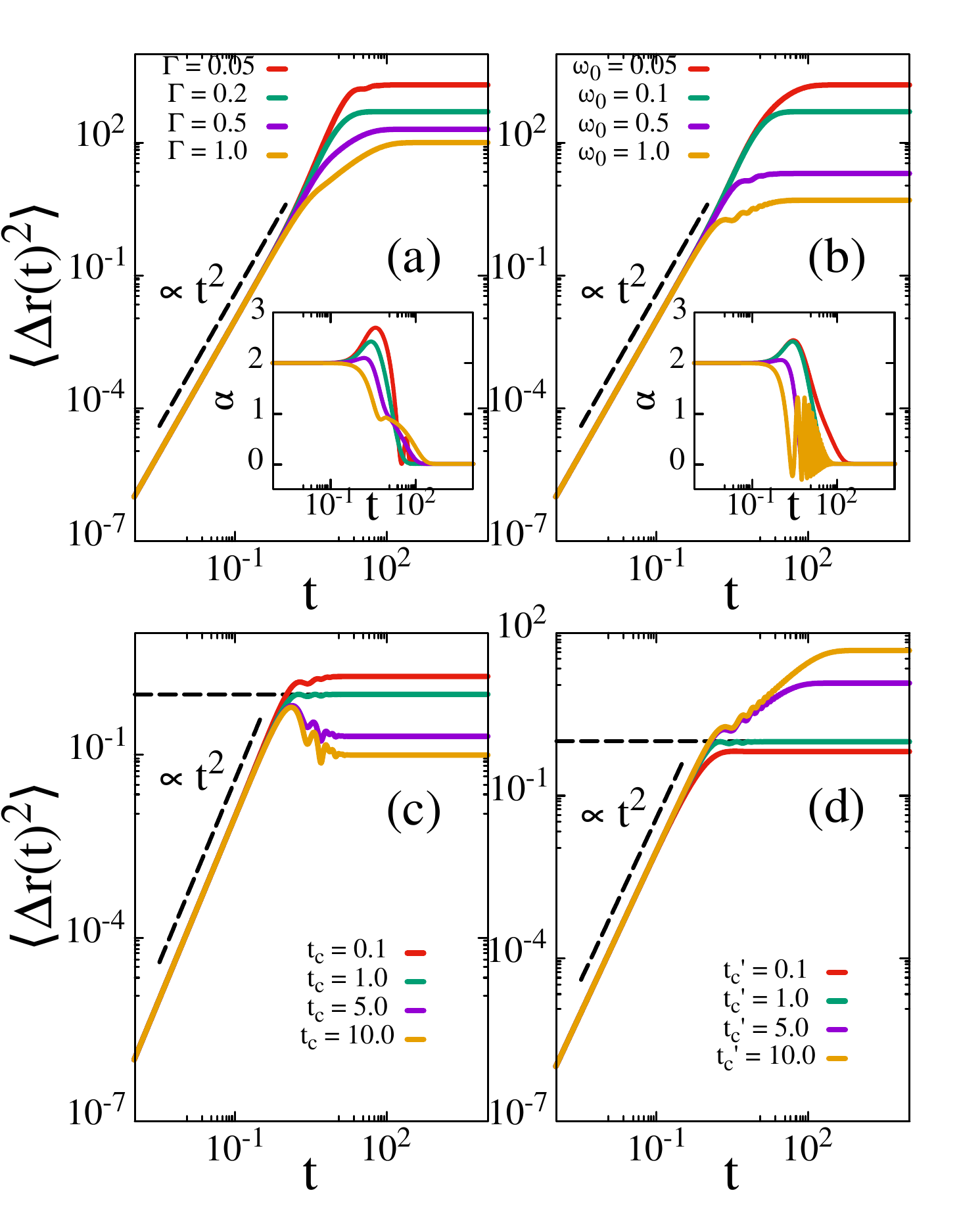}
    \caption{The MSD [Eq.~\eqref{eq:MSD-HT-no-mag}] as a function of $t$ for different values of $\Gamma$ in (a) for $t_c = t_c' = 1.0$ and $\omega_0 = 0.1$, for different values of $\omega_0$ in (b) for $t_c = t_c' = 1.0$ and $\Gamma = 0.2$, for different values of $t_c$ in (c) for a fixed $t_c'=1$, and for different values of $t_c'$ in (d) for a fixed $t_{c}=1$, respectively. The insets of (a) and (b) show the exponent $\alpha$ as a function of $t$ corresponding to the respective MSD plots. The dashed horizontal lines in (c) and (d) correspond to the steady state equilibrium value of MSD. The other common parameters are $D = m = 1.$}.
    \label{fig:MSD_HT_nomag}
\end{figure}

\begin{figure}
    \centering
    \includegraphics[scale=0.35]{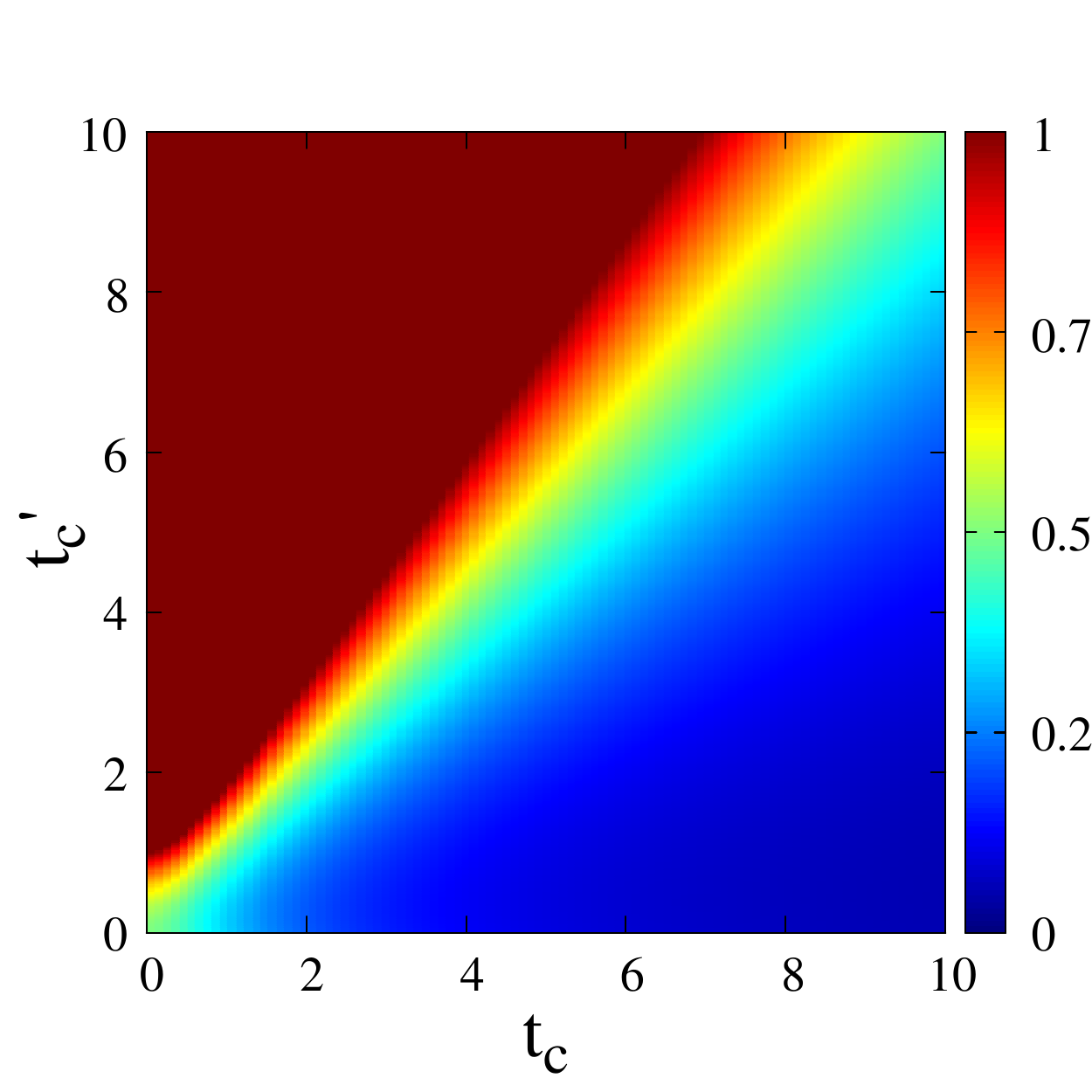}
    \caption{The 2d plot of steady state MSD $\langle \Delta r^2\rangle_s$ [Eq.~\eqref{eq:msd_st_eq}] as a function of both $t_c$ and $t_c'$. The other common parameters are $\Gamma=1.0$, $m = 1$, $D = 1$ and $\omega_0=1.0$.  }
    \label{fig:TE_MSD_HT}
\end{figure}

\begin{figure*}
    \centering
    \includegraphics[width=0.75\linewidth]{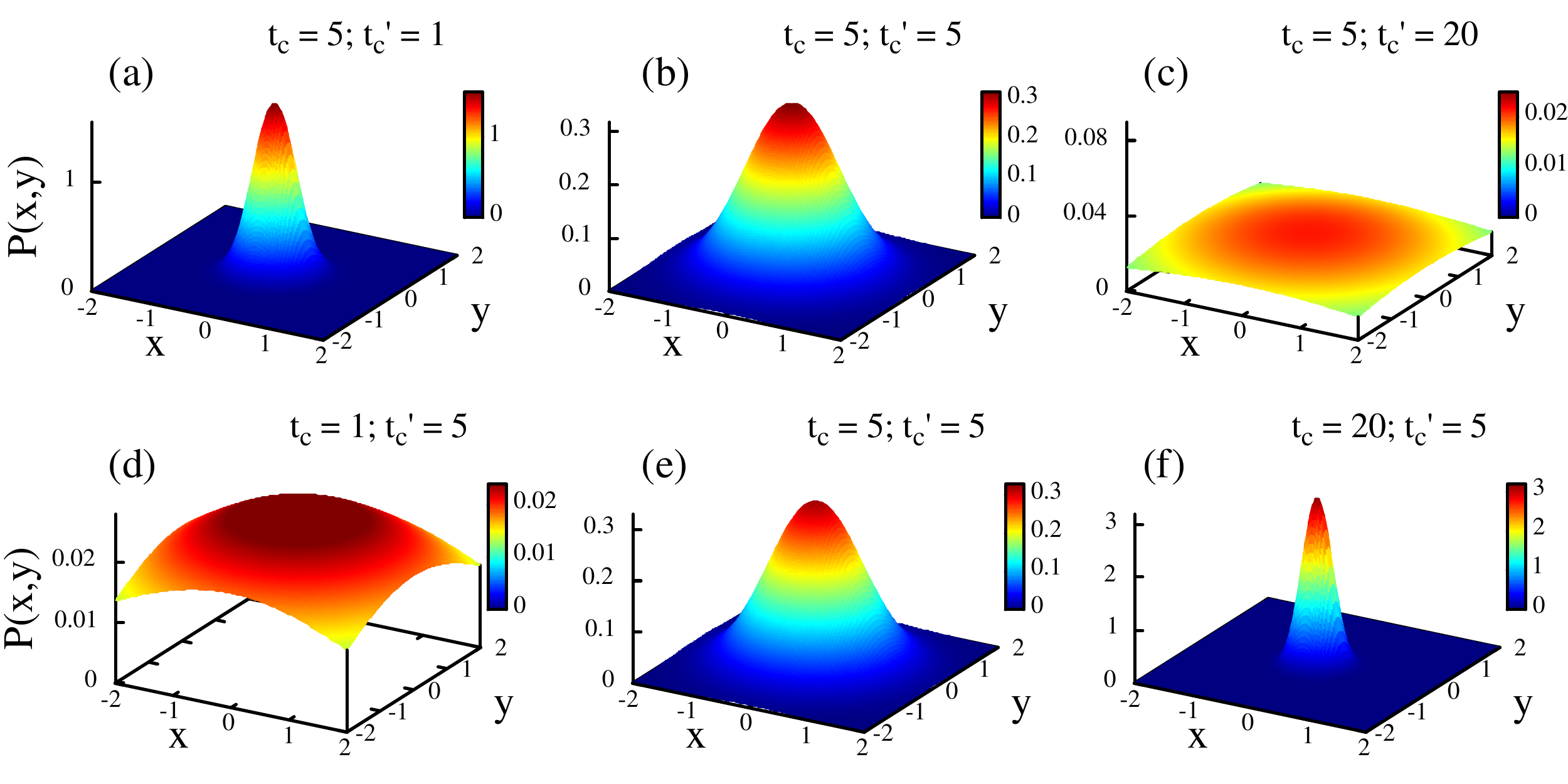}
    \caption{The steady state position probability distribution [Eq.~\eqref{eq:pdf_Pxy}] for different values of $t_c'$ fixing $t_c(=5.0)$ are shown in (a)-(c) and for different values of $t_c$ for a fixed $t_c'(=5)$ is shown in (d)-(f). Other parameters are $m = \Gamma = \omega_0 = D = 1$.}
    \label{fig:PD_nomag}
\end{figure*}
In Fig.~\ref{fig:MSD_HT_nomag}, we present $\langle \Delta {\bf r}^2(t) \rangle$ as a function of $t$ for different values of $\Gamma$ in Fig.~\ref{fig:MSD_HT_nomag}(a), $\omega_{0}$ in Fig.~\ref{fig:MSD_HT_nomag} (b), $t_{c}$ in Fig.~\ref{fig:MSD_HT_nomag}(c), and $t_c'$ in Fig.~\ref{fig:MSD_HT_nomag}(d), respectively. In order to explore different time regimes of motion, we introduce a parameter $\alpha$ such that $\langle \Delta {\bf r}^2(t) \rangle \propto t^\alpha$, i.e.,
\begin{equation}
    \alpha = \frac{\partial \log \langle \Delta {\bf r}^2(t) \rangle}{\partial \log t}.
\end{equation}
The insets of Fig.~\ref{fig:MSD_HT_nomag}(a) and (b) show the variation of the exponent $\alpha$ with $t$. The initial transient time MSD is always found to be ballistic (since $\langle \Delta {\bf r}^2(t) \rangle \propto t^2$ or $\alpha=2$) and the steady state or long time regime is always non-diffusive (since $\langle \Delta {\bf r}^2(t) \rangle$ is independent of $t$ or $\alpha=0$). However, the intermediate time regimes of MSD is found to be oscillatory for the parameters that fall in the oscillatory regime of parameter space. In Fig.~\ref{fig:MSD_HT_nomag}(a), the variation of MSD as a function of $t$ for low and high values of $\Gamma$ ($\Gamma=0.05, 0.5, 1.0$) show intermediate time oscillations and that is why the exponent $\alpha$ as a function of $t$ shows intermediate time oscillations for $\Gamma=0.05, 0.5$, and $1.0$, respectively, whereas the MSD is non-oscillatory for $\Gamma=0.2$. With increase in $\omega_{0}$ value in Fig.~\ref{fig:MSD_HT_nomag} (b), the system makes a transitions from no-oscillatory phase to oscillatory phase and results intermediate time oscillations in MSD for larger values of $\omega_0$ (for $\omega_{0}=0.5$ and $\omega_{0}=1$). The same can also be confirmed from the inset of Fig.~\ref{fig:MSD_HT_nomag} (b) which shows the intermediate time oscillatory behaviour of $\alpha$ for $\omega_0=0.5$ and $\omega_{0}=1$. Figures~\ref{fig:MSD_HT_nomag} (c) and (d) show the variation of MSD as a function of $t$ for different values of $t_c$ and $t_c'$, respectively. With an increase in $t_c$ values, the steady state MSD decreases (see Fig.~\ref{fig:MSD_HT_nomag} (c)) and it increases with an increase in $t_c'$ values (see Fig.~\ref{fig:MSD_HT_nomag} (d)). At the same time, the initial ballistic regime of MSD gets reduced with increase in $t_{c}$ value and increases with increase in $t_c'$ values.

We exactly evaluate the steady state MSD and probability distribution function using steady state correlation matrix method formalism \cite{caprini2021inertial,van2007stochastic}. Because of the linear dynamics of the model and Gaussian nature of noise, the steady state probability distribution is Gaussian. In order for evaluating these two quantities, we introduce the correlation matrix $\Xi$ with components $\chi_i$, $\chi_j$ such that
\begin{equation}
    [\Xi_{i,j}] = \langle \chi_i \chi_j \rangle - \langle \chi_i \rangle \langle \chi_j \rangle
\end{equation}
and the distribution function $\mathcal{P}({\boldsymbol{\chi}})$ as
\begin{equation}
      \mathcal{P}({\boldsymbol{\chi}}) = \frac{1}{(2\pi)^4 \sqrt{Det(\Xi)}} e^{\left[{-\frac{1}{2}\left( \boldsymbol{\chi}^T\right)  \Xi^{-1}\left(\boldsymbol{\chi} \right)}\right]}.
      \label{eq:total_pdf}
\end{equation}

Here, $\boldsymbol{\chi}$ is the column vector whose elements $\chi_i \in \left( x\ y\ v_x\ v_y\ w_x\ w_y\ \xi_x\ \xi_y \right)$ with $v_x$ and $v_y$ being the $x$ and $y$ components of velocity. The variables $w_x$ and $w_y$ are the $x$ and $y$ components of the vector ${\bf w}$ given by
\begin{equation}
    {\bf w} = \Gamma \int_{0}^{t} \lambda\left(t-t'\right) {\bf \dot{r}}(t')\, dt'.
\end{equation}

Introduction of ${\bf w}$ splits the non-Markovian model dynamics [Eq.~\eqref{eq:maineqmotion-vector}] into a set of Markovian equations as follows 
\begin{align}
    {\bf \dot{r}} &= {\bf v} \\
    {\bf \dot{v}} &= -{\bf w} - \omega_0^2 {\bf r}+\frac{\sqrt{D}}{m} \boldsymbol{\xi} \\
    {\bf \dot{w}} &= -\frac{{\bf w}}{t_c'} + \frac{\Gamma}{t_c'} {\bf v} \\
    {\bf \dot{\boldsymbol{\xi}}} &= -\frac{\boldsymbol{\xi}}{t_c} + \frac{\boldsymbol{\eta}}{t_c}.
\end{align}
The above set of equations can be expressed as
\begin{equation}
    \dot{\boldsymbol{\chi}} = A\boldsymbol{\chi} + B \boldsymbol{\eta'}
    \label{eq:dynamics-matrix}
\end{equation}

Here, the matrices $A$, $B$ and $\boldsymbol{\eta'}$ are given by
\begin{equation}
    A = \begin{pmatrix}
        0 & 0 & 1 & 0 & 0 & 0 & 0 & 0 \\
        0 & 0 & 0 & 1 & 0 & 0 & 0 & 0 \\
        -\omega_0^2 & 0 & 0 & 0 & -1 & 0 & \frac{\sqrt{D}}{m} & 0 \\ 
        0 & -\omega_0^2 & 0 & 0 & 0 & -1 & 0 & \frac{\sqrt{D}}{m} \\
        0 & 0 & \frac{\Gamma}{t_c'} & 0 & \frac{-1}{t_c'} & 0 & 0 & 0 \\
        0 & 0 & 0 & \frac{\Gamma}{t_c'} & 0 & \frac{-1}{t_c'} & 0 & 0 \\
        0 & 0 & 0 & 0 & 0 & 0 & \frac{-1}{t_c} & 0 \\
        0 & 0 & 0 & 0 & 0 & 0 & 0 & \frac{-1}{t_c}
    \end{pmatrix},
\end{equation}
\begin{equation}
    B = \begin{pmatrix}
        0 & 0 & 0 & 0 & 0 & 0 & 0 & 0 \\
        0 & 0 & 0 & 0 & 0 & 0 & 0 & 0 \\
        0 & 0 & 0 & 0 & 0 & 0 & 0 & 0 \\
        0 & 0 & 0 & 0 & 0 & 0 & 0 & 0 \\
        0 & 0 & 0 & 0 & 0 & 0 & 0 & 0 \\
        0 & 0 & 0 & 0 & 0 & 0 & 0 & 0 \\
        0 & 0 & 0 & 0 & 0 & 0 & \frac{1}{t_c} & 0 \\
        0 & 0 & 0 & 0 & 0 & 0 & 0 & \frac{1}{t_c} \\
    \end{pmatrix},
\end{equation}
and 
\begin{equation}
    \boldsymbol{\eta'} = (0\ 0\ 0\ 0\ 0\ 0\ \eta_x\ \eta_y)^T.
\end{equation}

As per the correlation matrix formalism, the correlation matrix $\Xi$ can be shown to satisfy the equation
\begin{equation}
    A\cdot \Xi + \Xi \cdot A^T + B\ B^T = 0.
    \label{eq:Xi_relation}
\end{equation}

Finally, solving the above Eq.~\eqref{eq:Xi_relation}, the MSD at steady state  $\langle \Delta{\bf r}^2 \rangle_s$ can be calculated as
\begin{equation}
    \begin{split}
     \langle \Delta{\bf r}^2 \rangle_s &= \Xi_{1,1} + \Xi_{2,2}\\
     & = \frac{D\left[t_c + t_c' + t_c^2\Gamma + t_c'^2(t_c + t_c')\omega_0^2\right]}{m^2\Gamma \omega_0^2\left[t_c' + t_c(1 + t_c\Gamma + t_c(t_c + t_c')\omega_0^2)\right]} \\
     & = \frac{D}{\Gamma m^2 \omega_0^2   }+\frac{D \left(t_c'-t_c\right) \left(t_c'+t_c\right){}^2}{\Gamma  m^2 \left(\Gamma  t_c^2+\omega _0^2 t_c^2 t_c'+\omega _0^2 t_c^3+t_c'+t_c\right)}.
    \end{split}
     \label{eq:msd_st_eq}
\end{equation}

From the above equation, it is confirmed that for $t_c = t_c'$ limit, $\langle \Delta{\bf r}^2 \rangle_s$ reduces to $\frac{D}{\Gamma m^2 \omega_0^2}$. For $D=2 \gamma k_{B} T$, it approaches the equilibrium value given by $\langle \Delta{\bf r}^2 \rangle_{eq} = \frac{2 k_BT}{k}$. Similarly, in $\omega_{0} \to \infty$ limit, $\langle \Delta{\bf r}^2 \rangle_s$ approaches the equilibrium value for $D=2 \gamma k_{B} T$ even when $t_c \neq t_c'$. However, in this limit, the system doesn't approach equilibrium. The system approaches thermal equilibrium for $D=2\gamma k_{B} T$ in $t_c=t_c'$ limit and the equilibrium value follows the equipartition theorem
\begin{equation}
    \frac{1}{2} k\langle \Delta{\bf r}^2 \rangle_{eq} = k_BT.
\end{equation}

The first term of Eq.~\eqref{eq:msd_st_eq} is independent of $t_c$ and $t_c'$. With increase in $t_c'$ value, the second term of Eq.~\eqref{eq:msd_st_eq} becomes dominant and hence in $t_c' \to \infty$ limit, this term diverges. As a result, the steady state MSD diverges, i.e.,
\begin{equation}
    \lim\limits_{t_c'\to \infty} \langle \Delta{\bf r}^2 \rangle_s = \infty.
\end{equation}
Similarly, with increase in $t_{c}$ value, the second term of Eq.~\eqref{eq:msd_st_eq} becomes negative, as a result the $\langle \Delta{\bf r}^2 \rangle_s$ decreases. In $t_c \to \infty$ limit, the second term of Eq.~\eqref{eq:msd_st_eq} becomes $-\frac{D}{\Gamma m^2 \omega_0^2}$, which is exactly equal to the negative of the first term. Hence, in this limit, the steady state MSD vanishes, i.e.,
\begin{equation}
    \lim\limits_{t_c\to \infty} \langle \Delta{\bf r}^2 \rangle_{s} = 0.
    \label{eq:msd_st_tc_infty}
\end{equation}

These results are summarized in Fig.~\ref{fig:TE_MSD_HT}, where we present the 2d plot of $\langle \Delta{\bf r}^2 \rangle_{s}$ as a function of both $t_c$ and $t_c'$. It is observed that for a fixed value of $t_c$, $\langle \Delta{\bf r}^2 \rangle_{s}$ increases with increase in $t_c'$ and for sufficiently large value of $t_c'$, $\langle \Delta{\bf r}^2 \rangle_{s}$ becomes infinitely large, indicating the escape of the particle without approaching steady state. This observation suggests that persistence of sufficiently long duration of memory in the medium provides a kind of elastic bound to the particle that overcomes the harmonic confinement as a result of which the particle can escape out of the potential and takes infinitely long time to come back the mean position of the well or approach the steady state.  On the other hand, for a fixed value of $t_c'$, $\langle \Delta{\bf r}^2 \rangle_{s}$ decreases with increase in $t_c$ value and becomes zero for infinitely large value of $t_c$. This fact indicates the trapping of the particle for the presence of infinitely long duration of activity in the medium. This is the genuine feature of an active particle confined by a finite potential~\cite{muhsin2023inertialactive}. 

Now, substituting the solution $\Xi$ from Eq.~\eqref{eq:Xi_relation} and integrating it over all other variables, we obtain the steady state joint probability distribution function $P(x, y)$ as 
\begin{equation}
    P(x, y)=\frac{1}{2\pi\sigma^2} \exp{\left(-\frac{ x^2 + y^2}{2\sigma^2}\right)},
    \label{eq:pdf_Pxy}
\end{equation}
with the variance $\sigma^2$ given by
\begin{equation}
    \sigma^2 = \frac{D\left[t_c + t_c' + t_c^2\Gamma + t_c'^2(t_c + t_c')\omega_0^2\right]}{2 m^2\Gamma \omega_0^2\left[t_c' + t_c(1 + t_c\Gamma + t_c(t_c + t_c')\omega_0^2)\right]}.
    \label{eq:Pxy_variance}
\end{equation}

In $t_c' \to 0$ limit, the distribution is still Gaussian with the variance [Eq.~\ref{eq:Pxy_variance}] as
\begin{equation}
    \lim_{t_c'\to 0} \sigma^2 = \frac{D(1 + t_c\Gamma)}{2m^2\Gamma\omega_0^2\left( 1 + t_c\Gamma + t_c^2\omega_0^2 \right)}.
\end{equation}

In Figs.~\ref{fig:PD_nomag} (a)-(c), we plot the steady state probability distribution $P(x,y)$ for different values of $t_c'$ and for a fixed value of $t_{c}=5$. For all $t_c'$ values, the distribution is Gaussian centered at the origin of the potential. As $t_c'$ increases, $\sigma^2$ becomes an increasing function of $t_c'$. Hence the variance or width of the distribution increases and the distribution spreads out [see Figs.~\ref{fig:PD_nomag} (a)-(c)]. At the same time, the peak of the distribution suppresses. This implies that with increase in memory time scale, the probability of finding the particle at the mean position of the well decreases and the probability of finding the particle at larger distances increases. Finally, in $t_c' \to \infty$ limit, $P(x,y) \to 0$ and the distribution becomes flat [Fig.~\ref{fig:PD_nomag} (c)]. The vanishing of the position distribution function in $t_c' \to \infty$ limit represents the escape of the particle out of the potential. This observation supports the enhancement of particle trajectories [Fig.~\ref{fig:Traj_g10_HT_nomag_tcp}] and the enhancement of steady state MSD [Fig.~\ref{fig:TE_MSD_HT}] with increase of the persistent duration of memory. 

Similarly, in Fig.~\ref{fig:PD_nomag} (d)-(e), we show the plot of $P(x,y)$ for different values of $t_c$ and for a fixed $t_c'$ value. Initially, in the $t_c\to 0$ limit,  $P(x,y$ is Gaussian [Eq.~\ref{eq:pdf_Pxy}] with $\sigma^2$ given by
\begin{equation}
    \lim_{t_c\to 0} \sigma^2 = \frac{D(1 + t_c'^2\omega_0^2)}{2m^2\Gamma\omega_0^2}.
\end{equation}
 As $t_c$ is increased, $\sigma^2$ becomes decreasing function of $t_{c}$ and hence the variance (or width) of the distribution decreases [see Figs.~\ref{fig:PD_nomag} (d)-(e)]. Simultaneously,  the peak of the distribution increases and the distribution becomes narrow. Finally, in $t_c\to\infty$ limit, $\sigma^2$ approaches zero value and hence the probability distribution becomes a delta function since
    \begin{equation}
\begin{split}
    \lim_{t_c\to \infty}P(x,y) &= \lim_{\sigma\to 0} \frac{1}{2\pi\sigma^2} \exp{\left(-\frac{ x^2 + y^2}{2\sigma^2}\right)}\\
    &= \delta(x) \delta(y).
\end{split}
\end{equation}

This suggests that with increase in $t_c$ value, the chances of finding the particle at the mean position of the well increases, confirming the trapping of the particle for sufficiently long duration of activity in the medium. These results are complemented with the suppression of particle trajectories in Fig.~\ref{fig:Traj_g10_HT_nomag_tc} and reduction of steady state MSD in Fig.~\ref{fig:TE_MSD_HT} with increase in $t_{c}$ values. 

\section{SUMMARY AND CONCLUSIONS}\label{sec:summary}
In summary, we have explored the self propulsion of a harmonically confined inertial active Ornstein Uhlenbeck particle in a non-Newtonian environment, characterized by viscoelastic suspension. We model the dynamics using the generalized Langevin equation of motion. By solving this non-Markovian model, we demonstrate a phase diagram distinguishing two separate regimes in $t_c' \omega_{0}-t_c' \Gamma$ parameter space associated with the oscillatory and non-oscillatory behavior of the solution. From the exact solution of the dynamics and from the simulation results, both transient and steady state properties of motion are investigated. The simulated particle trajectories reveal that the particle exhibits rotational motion in the oscillatory regime of the parameter space. The rotational trajectories keep on getting enhanced with increase in persistent duration of memory and gets suppressed with increase in persistent duration of activity in the medium. This observation suggests that with increase in memory time scale, the elastic influence exerted by the environment on the particle dominates over the harmonic bound of the potential. As a consequence, the particle can come out of the potential without approaching steady state. This result is further supported by the enhancement of steady state mean square displacement and uniform spreading of position distribution function through out the space with increase in the memory time scale. Similarly, the suppression of particle trajectories as an increasing function of activity time is also complemented with the reduction of steady state mean square displacement and narrowing of the probability distribution function with increase in activity time scale. 
\section{Acknowledgement}
We thank the 8th statphysics community meeting (ICTS/ISPCM2023/02), during which some parts of the work were done. MS acknowledges the start-up grant from UGC, state plan fund from the University of Kerala, and SERB-SURE grant (SUR/2022/000377) from DST, Govt. of India for financial support. MM acknowledges SERB international travel grant (ITS/2023/002740) from DST, Govt. of India for financial support. 

\end{document}